\documentclass[10pt]{article}

\usepackage{url, graphicx, color, varioref, amsfonts, longtable, float}

\newcommand{\qed}{\nobreak \ifvmode \relax \else
      \ifdim\lastskip<1.5em \hskip-\lastskip
      \hskip1.5em plus0em minus0.5em \fi \nobreak
      \vrule height0.75em width0.5em depth0.25em\fi}

\title{Automated Method for Building $CNOT$ Based Quantum Circuits for Boolean Functions}

\author{
Ahmed Younes \and Julian Miller\\
\and\\
School of Computer Science\\ 
The University of Birmingham\\ 
Birmingham\\
B15 2TT\\
United Kingdom\\
\{A.Younes , J.Miller\}@cs.bham.ac.uk \\
}

\begin{document}
\maketitle
\begin{abstract}
In this paper we discuss an efficient technique that can 
implement any given Boolean function as a quantum circuit. 
The method converts a truth table of a Boolean function to 
the corresponding quantum circuit using a minimal number of 
auxiliary qubits. We give examples of some circuits 
synthesized with this technique. A direct result that 
follows from the technique is a new way to convert any 
classical digital circuit to its classical reversible form.
\end{abstract}

\section{Introduction}       
Implementing Boolean functions on quantum computers is an essential aim,
in the exploration of the benefits, which may be gained from systems 
operating by quantum rules. It is important to find the corresponding quantum 
circuits, which can carry out the operations we use to implement on our 
conventional computers. On classical computers, a circuit can be built for 
any Boolean function using AND, OR and NOT gates. This set of gates cannot, in general 
be used to build quantum circuits because the operations are not reversible \cite{toffoli80}. A 
corresponding set of reversible gates must be used to build a quantum 
circuit for any Boolean operation. In classical computer science, many clever methods have been used to 
obtain more efficient digital circuits \cite{old:bok} for a given Boolean function. 
Recently, there have been efforts to find an automatic way to create efficient quantum circuits implementing 
Boolean functions. It is shown that \cite{elementary-gates} any unitary gate can be represented 
as a composition of simpler gates but it is not necessarily the most 
efficient circuit for this operation. A method proposed in \cite{practmethod} used a 
modified version of \textit{Karnaugh maps}\cite{old:bok} and depends on a clever choice of certain minterm gates 
to be used in minimization process, however it appears that this method has poor scalability. Another work \cite{transrules}, 
includes a very useful set of 
transformations for quantum Boolean circuits and proposes a method for 
building quantum circuits for Boolean functions by using extra 
auxiliary qubits, however, this will increase the number of qubits to be used in the 
final circuits.

\begin{figure}[htbp]
\centerline{\includegraphics[width=3.937cm,height=3.937cm]{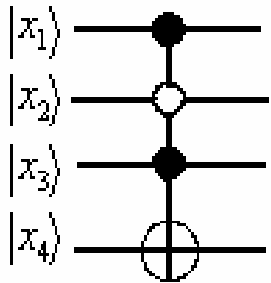}}
\caption{$CNOT$ gate.}
\label{BQCfig1}
\end{figure}

In our construction for building quantum circuits for Boolean functions, we 
will use only one auxiliary qubit; which we initially set to zero, to hold the result of the Boolean function,
together with $CNOT$ based transformations (gates) which work as follows \cite{transrules}: $CNOT\left( {C|t} \right)$ is a gate where the 
target qubit $t$ is controlled by a set of qubits $C$ such that $t \notin C$, the 
state of the qubit $t$ will be flipped from $\left| 0 \right\rangle $ to 
$\left| 1 \right\rangle $ or from $\left| 1 \right\rangle $ to $\left| 0 
\right\rangle $ if and only if the conditions stated by the $CNOT$ gate is 
evaluated to true. The condition that a certain qubit evaluates to true depends on whether the 
state of the qubit is $\left| 0 \right\rangle $ 
(cond-0; $\delta $=1) or $\left| 1 \right\rangle $ (cond-1; $\delta $=0) 
according to the condition being set, where $\delta $ is a Boolean parameter 
that will be used in the Boolean algebraic expressions to indicate the condition 
being set on the qubit; i.e. the new state of target qubit $t$ is the result of $XOR$-ing 
the old state of $t$ with the $AND$-ing of the states of the control qubits $C$ (under the
condition being set on each control qubit). For example, consider the $CNOT$ gate shown in Fig.\ref{BQCfig1}, it 
can be represented as $CNOT\left( {\left\{ {x_1 ,\overline {x_2 } ,x_3 } 
\right\}\vert x_4 } \right)$, where $ \circ $ and $\bullet $ mean that the 
condition on the qubit will evaluate to true if and only if the state of 
that qubit is $\left| 0 \right\rangle $ (cond-0) and $\left| 1 \right\rangle 
$ (cond-1) respectively, while $ \oplus $ denotes the target qubit which 
will be flipped if and only if all the conditions set on the control qubits 
being evaluated to true. This means that the state of the qubit $x_{4}$ will be 
flipped if and only if $x_{1}=x_{3}=\left| 1 \right\rangle $ and 
$x_{2}=\left| 0 \right\rangle $. In general, the target qubit in a 4--qubit 
gate will be changed according to the operation $x_4 \to x_4 \oplus \left( 
{x_1 \oplus \delta _1 } \right)\left( {x_2 \oplus \delta _2 } \right)\left( 
{x_3 \oplus \delta _3 } \right)$, now to represent the gate shown in Fig.\ref{BQCfig1}, 
we will set $\delta _{1 }=\delta _{3}= 0$ and $\delta _{2}= 1$, 
so the operation for this gate on $x_{4}$ will be $x_4 \to x_4 \oplus x_1 
\overline {x_2 } x_3 $,

\begin{figure} [htbp]
\centerline{\includegraphics [width=7.58444cm,height=3.77444cm]{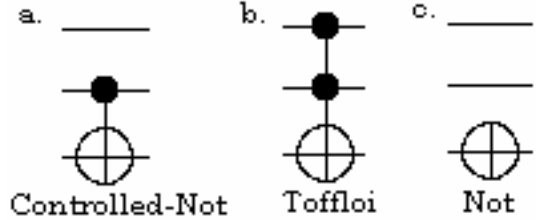}} 
\vspace*{13pt}
\caption{Special cases of the general $CNOT$ gate.}
\label{BQCfig2}
\end{figure}

Some special cases of the general $CNOT$ gate have their own names, $CNOT$ gate with 
one control qubit with cond-1 is called Controlled-$Not$ gate; Fig.\ref{BQCfig2}(a), $CNOT$ gate with two 
control qubits both with cond-1 is called \textit{Toffoli} gate; Fig.\ref{BQCfig2}(b), and $CNOT$ gate with 
no control qubits at all is called $NOT$ gate; Fig.\ref{BQCfig2}(c), where $C$ will be an empty 
set (C = $\Phi )$, we will refer to this case as $CNOT\left( x_{i} \right)$ where $x_{i}$  is 
the qubit which will be unconditionally flipped.

\section{Quantum Boolean Function}
\noindent

A Boolean function, $F$, is a function that takes $n$ Boolean variables as inputs and 
gives one Boolean variable as output,

\begin{equation}
\label{BQCeq1}
F(x_1 ,x_2 ,\ldots ,x_n ) \to \left\{ {0,1} \right\},	x_i \in \left\{ {0,1} 
\right\} \quad  
\end{equation}

To represent a Boolean function of $n-1$ inputs, a quantum circuit with $n$ qubits will be 
used where the extra qubit will be initialised with value $0$, this will then carry the 
result of the Boolean function at the end of the computation. Any Boolean 
function can be represented by a truth table, In order to be reversible; 
the truth table must have $n$ inputs and $n$ outputs. For example: Consider the 
Boolean function $F(x_1 ,x_2 ,x_3 ) = \overline {x_1 } + x_2 x_3 $, 
classically it's truth table is represented as shown in Table.\ref{BQCtab1} and for 
quantum computing purposes, the representation will be as shown in Table.\ref{BQCtab2}.

\begin{table}[H]
\begin{center}
\begin{tabular}{|c|c|c|c|}
\hline
$x_{1}$  & $x_{2}$& 
$x_{3}$   & 
F  \\
\hline
0  & 0  & 0  & 1  \\ \hline
0  & 0 & 1   & 1  \\ \hline
0  & 1  & 0  & 1   \\ \hline
0  & 1  & 1  & 1   \\ \hline
1  & 0  & 0  & 0   \\ \hline
1  & 0  & 1  & 0   \\ \hline
1  & 1  & 0  & 0   \\ \hline
1  & 1  & 1  & 1   \\ \hline
\end{tabular}
\caption{Classical representation of the truth table for $F(x_1 ,x_2 ,x_3 ) 
= \overline {x_1 } + x_2 x_3 $.}
\label{BQCtab1}
\end{center}
\end{table}

\begin{table}[H]
\begin{center}
\begin{tabular}{|c|c|c|c|c|c|c|c|}
\hline

$x_{1}$   & $x_{2}$   & $x_{3}$   & $F_{ini}$   & $x_{1}$   & $x_{2}$   & $x_{3}$   & $F_{fin}$    \\\hline
0  & 0  & 0  & 0  & 0  & 0  & 0  & 1   \\\hline
0  & 0  & 1  & 0  & 0  & 0  & 1  & 1   \\\hline
0  & 1  & 0  & 0  & 0  & 1  & 0  & 1   \\\hline
0  & 1  & 1  & 0  & 0  & 1  & 1  & 1   \\\hline
1  & 0  & 0  & 0  & 1  & 0  & 0  & 0   \\\hline
1  & 0  & 1  & 0  & 1  & 0  & 1  & 0   \\\hline
1  & 1  & 0  & 0  & 1  & 1  & 0  & 0   \\\hline
1  & 1  & 1  & 0  & 1  & 1  & 1  & 1   \\\hline
\end{tabular}
\caption{Quantum computing version of the truth table for $F(x_1 ,x_2 ,x_3 ) 
= \overline {x_1 } + x_2 x_3 $.}
\label{BQCtab2}
\end{center}
\end{table}

From the second representation, we can see that the $F_{ini}$ will be 
flipped only if the result of the function $F$ is 1, $F(x_1 ,x_2 ,x_3 ) = 1$.

\section{Automatic Construction of Quantum Boolean Circuits }
\subsection*{\bf Stage 1:}

A quantum Boolean circuit $U$ of size $m$ over $n$ qubit quantum system with 
qubits $\left| {x_1 } \right\rangle ,\left| {x_2 } \right\rangle ,\ldots 
,\left| {x_n } \right\rangle $ can be represented as a sequence of $CNOT$ gates \cite{transrules}, 

\begin{equation}
U = CNOT\left( {C_1 \vert t_1 } \right)\ldots CNOT\left( {C_i \vert t_i } 
\right)\ldots CNOT\left( {C_m \vert t_m } \right), \quad 
\label{BQCeq2}
\end{equation}

\noindent
where $t_i \in \left\{ {x_1 ,\ldots ,x_n } \right\};\,\,C_i \subset \left\{ 
{x_1 ,\ldots ,x_n } \right\};\,t_i \notin C_i$.

Using the modified truth table, we will choose $CNOT\left( {C_i |t_i } \right)$ according to the following steps: 

\begin{enumerate}
\item[1.] Select the input configurations from the truth table where $F_{fin}$ is 1.

\item[2.] Add a \textit{single} $CNOT$ gate for \textit{every} selected configuration taking the $F_{ini}$ as 
the target qubit.

\item[3.] Set the condition on the control qubit for gates being added according to it's value in the 
configuration from the truth table, i.e. the qubit with value 0 in the truth 
table will be set to cond-0 in the corresponding $CNOT$ gate and the qubit with 
value 1 will be set to cond-1 in the corresponding $CNOT$ gate. 

\item[4.]For input configurations where $F_{fin}$ is 0, we will not add any gates (as 
if we are applying identities on them).
\end{enumerate}

\begin{figure} [htbp]
\centerline{\includegraphics[width=8.382cm,height=4.953cm]{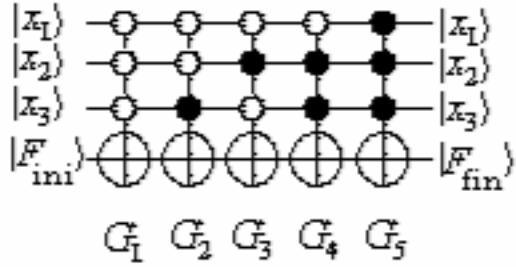}} 
\vspace*{13pt}
\caption{\label{BQCfig3}Initial quantum circuit for $F(x_1 ,x_2 ,x_3 ) = \overline {x_1 } + x_2 x_3 $.}
\end{figure}

For example, according to the truth table shown in Table.\ref{BQCtab2}, we will select 
only the configurations with $F_{fin}=1$ as shown in Table.\ref{BQCtab3} and construct 
the corresponding quantum circuit as shown in Fig.\ref{BQCfig3}.

\begin{table}[hb]
\begin{center}
\begin{tabular}{|c|c|c|c|c|}
\hline
 	& $x_{1}$   & $x_{2}$   & $x_{3}$   & $F_{fin}$    \\\hline
$G_1$  & 0  & 0  & 0  & 1   \\\hline
$G_2$  & 0  & 0  & 1  & 1   \\\hline
$G_3$  & 0  & 1  & 0  & 1   \\\hline
$G_4$  & 0  & 1  & 1  & 1   \\\hline
$G_5$  & 1  & 1  & 1  & 1   \\\hline
\end{tabular}
\caption{Input configurations where $F(x_1 ,x_2 ,x_3 ) = \overline {x_1 } + x_2 x_3 = 1.$}
\label{BQCtab3}
\end{center}
\end{table}

The maximum number of \textit{CNOT} gates we can add in this stage will be up to 2$^{n - 
1}$ \textit{CNOT} gate where $n$ is the number of qubits in the quantum system.

\subsection*{\bf Stage 2:}

In the following transformations we will trace the operations being 
applied on the target qubit only, since no control qubits will be changed 
during the operations of the circuit. These circuit transformations are an extension and 
generalization of some of the equivalence between reversible circuits shown 
in \cite{revlogic}. We will apply this transformations on every $CNOT$ gate in the circuit we have,
which will expand the number of $CNOT$ gates in the circuit, after which we will apply
the Rule of Minimization on the whole circuit to get the final circuit, which
implements the Boolean finction.

Let $x_{i}$'s be the control qubit, $x_{n}$ be the target qubits and $\delta _i \in 
\left\{ {0,1} \right\}$ where i=1,2,\ldots ,$n$-1, the general operation to be 
applied on the target qubit is given by

\begin{equation}
\label{BQCeq9}
x_n \to x_n \oplus \left( {x_1 \oplus \delta _1 } \right)\left( {x_2 \oplus 
\delta _2 } \right)\ldots \left( {x_{n - 2} \oplus \delta _{n - 2} } \right)\left( {x_{n - 1} \oplus \delta _{n - 1} } \right) 
\quad 
\end{equation}

\noindent
Multiplying all terms we get the following transformation:

\begin{equation}
\label{BQCeq10}
\begin{array}{l}
 x_n \oplus \left( {x_1 \oplus \delta _1 } \right)\left( {x_2 \oplus \delta 
_2 } \right)\ldots \left( {x_{n - 2} \oplus \delta _{n - 2} } \right)\left( {x_{n - 1} \oplus \delta _{n - 1} } \right) \\ 
 	 = x_n \oplus x_1 x_2 \ldots x_{n - 1} \oplus x_1 x_2 \ldots x_{n - 2}\delta _{n - 
1} \oplus \ldots \oplus \delta _1 x_2 \ldots x_{n - 1} \oplus \ldots \oplus 
\delta _1 \delta _2 \ldots \delta _{n - 1} \\ 
\end{array} \quad 
\end{equation}

\subsection*{\bf Examples}
\begin{figure} [htbp]
\centerline{\includegraphics[width=6.526cm,height=4.480cm]{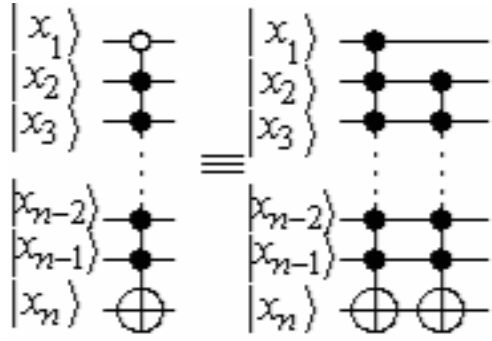}} 
\vspace*{13pt}
\caption{\label{BQCfig9}$n$ qubit gate with only $\delta _{1}$=1 and it's equivalent 
circuit.}
\end{figure}

{\it Example 1:} If one control qubit with cond-0. Let $\delta _{1}$ = 1 
as shown in Fig.\ref{BQCfig9}. The following two circuits are equivalent:

\begin{equation}
\begin{array}{l}
 CNOT(x_1 ).CNOT(\{ x_1 ,x_2 , \ldots ,x_{n - 1} \} |x_n ).CNOT(x_1 ){\rm  } \\ 
  = {\rm  }CNOT(\{ x_1 ,x_2 , \ldots ,x_{n - 1} \} |x_n ).{\rm  }CNOT(\{ x_2 ,{\rm  }x_3 ,{\rm  } \ldots ,{\rm  }x_{n - 1} \} |x_n ){\rm   } \\ 
 {\rm   } \\ 
 \end{array}
\label{BQCeq11_1}
\end{equation}

\subsubsection*{Proof:} 

From Eqn.\ref{BQCeq10}, putting $\delta _{1}$ = 1 and $\delta _i = 0;\,i = 2,\ldots ,n 
- 1$ to get Eqn.\ref{BQCeq11}. The L.H.S. of Eqn.\ref{BQCeq11} will represent L.H.S. circuit in 
Fig.\ref{BQCfig9}, and the R.H.S of Eqn.\ref{BQCeq11} will represent the R.H.S. circuit in Fig.\ref{BQCfig9}.

\begin{equation}
\label{BQCeq11}
x_n \oplus \overline {x_1 }\, x_2 \ldots x_{n - 1} = x_n \oplus x_1 x_2 \ldots 
x_{n - 1} \oplus x_2 \ldots x_{n - 1} \quad 
\end{equation}

\begin{figure} [htbp]
\centerline{\includegraphics[width=8.502cm,height=4.480cm]{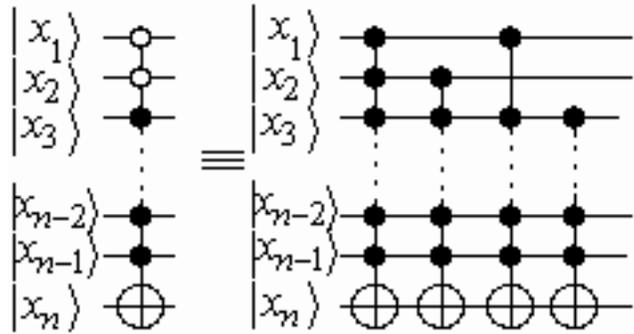}} 
\vspace*{13pt}
\caption{\label{BQCfig10}$n$--qubit gate with $\delta _{1}$=1 and $\delta _{2}$=1 and it's 
equivalent circuit.}
\end{figure}

{\it Example 2:} If two control qubit with cond-0. Let $\delta _{1}$ = 1 and 
$\delta _{2}$ = 1 as shown in Fig.\ref{BQCfig10}. The following two circuits are equivalent:

\begin{equation}
\begin{array}{l}
 CNOT(x_1 ).CNOT(x_2 ).CNOT(\{ x_1 ,x_2 , \ldots ,x_{n - 1} \} |x_n ).\\
 CNOT(x_2 ).CNOT(x_1 ) = CNOT(\{ x_1 ,x_2 , \ldots ,x_{n - 1} \} |x_n ).\\
 CNOT(\{ x_2 ,{\rm  }x_3 , \ldots ,x_{n - 1} \} |x_n ).CNOT(\{ x_1 ,x_3 , \ldots ,x_{n - 1} \} |x_n ).\\
 CNOT(\{ x_3 ,x_4 , \ldots ,x_{n - 1} \} |x_n )\\ 
 \end{array}
\label{BQCeq12}
 \end{equation}

\subsubsection*{Proof:}

From Eqn.\ref{BQCeq10}, putting $\delta _{1}$ = 1, $\delta _{2}$ = 1 and $\delta _i 
= 0;i = 3,\ldots ,n - 1$ to get Eqn.\ref{BQCeq13}. The L.H.S. of Eqn.\ref{BQCeq13} will represent 
the L.H.S. circuit in Fig.\ref{BQCfig10}, and the R.H.S of Eqn.\ref{BQCeq13} will represent the 
R.H.S. circuit in Fig.\ref{BQCfig10}.

\begin{equation}
\label{BQCeq13}
x_n \oplus \overline {x_1 }\,\, \overline {x_2 } \ldots x_{n - 1} = x_n \oplus 
x_1 x_2 \ldots x_{n - 1} \oplus x_2 x_3 \ldots x_{n - 1} \oplus x_1 x_3 
\ldots x_{n - 1} \oplus x_3 x_4 \ldots x_{n - 1} \quad
\end{equation}

\begin{figure} [htbp]
\centerline{\includegraphics[width=15.346cm,height=4.480cm]{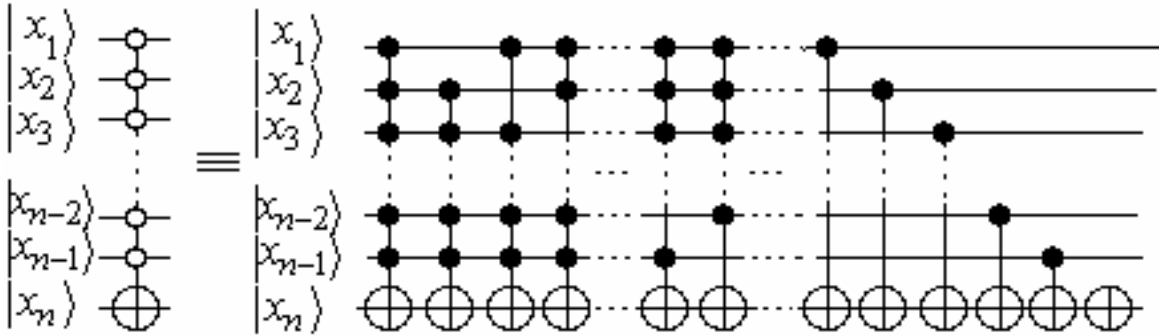}} 
\vspace*{13pt}
\caption{\label{BQCfig11}$n$--qubit gate with all $\delta _{i}$=1 and it's equivalent 
circuit.}
\end{figure}

{\it Example 3:} The last possible case where 
all $\delta $'s will be equal to 1 as shown in Fig.\ref{BQCfig11}. The following two circuits are equivalent:

\begin{equation}
\begin{array}{l}
 CNOT(x_1 ).CNOT(x_2 ) \ldots CNOT(x_{n - 1} ).CNOT(\{ x_1 ,x_2 , \ldots ,x_{n - 1} \} |x_n ). \\ 
 CNOT(x_{n - 1} ) \ldots CNOT(x_2 ).CNOT(x_1 ){\rm  } = {\rm  }CNOT(\{ x_1 ,x_2 , \ldots ,x_{n - 1} \} |x_n ). \\ 
 CNOT(\{ x_2 ,{\rm  }x_3 , \ldots ,{\rm  }x_{n - 1} \} |x_n ).CNOT(\{ x_1 ,{\rm  }x_3 , \ldots ,{\rm  }x_{n - 1} \} |x_n ).{\rm  } \\ 
 CNOT(\{ x_1 ,{\rm  }x_2 ,x_4 , \ldots ,{\rm  }x_{n - 1} \} |x_n ) \ldots CNOT(\{ x_1 ,{\rm  }x_2 , \ldots ,{\rm  }x_{n - 3} ,{\rm  }x_{n - 1} \} |x_n ). \\ 
 CNOT(\{ x_1 ,{\rm  }x_2 , \ldots ,{\rm  }x_{n - 2} \} |x_n ) \ldots CNOT(\{ x_1 \} |x_n ).{\rm  }CNOT(\{ x_2 \} |x_n ) \\ 
  \ldots CNOT(\{ x_{n - 1} \} |x_n ).CNOT(x_n ) \\ 
 \end{array}
\label{BQCeq14}
\end{equation}

\subsubsection*{Proof :}

From Eqn.\ref{BQCeq10}; putting all $\delta _{i}$ = 1, $i$=1,\ldots ,$n$-1 to get Eqn.\ref{BQCeq15}, 
The L.H.S. of Eqn.\ref{BQCeq15} represents the L.H.S. circuit in Fig.\ref{BQCfig11} and The R.H.S. 
of Eqn.\ref{BQCeq15} represents the R.H.S. circuit in Fig.\ref{BQCfig11}.

\begin{equation}
\label{BQCeq15}
 x_n \oplus \overline {x_1 }\,\, \overline {x_2 } \ldots \overline {x_{n - 1} } 
= \overline V , 
\end{equation} 
\noindent
where,
\[
\begin{array}{l}
 V = x_n \oplus x_1 x_2 \ldots x_{n - 1} \oplus x_2 x_3 \ldots x_{n - 1}\oplus x_1 x_3 \ldots x_{n - 1} \\
 \oplus x_1 x_2 x_4 \ldots x_{n - 1} \oplus \ldots \oplus x_1 x_2 \ldots x_{n - 3} x_{n - 1} \oplus x_1 x_2 \ldots x_{n - 2} \\
 \oplus x_1 \oplus x_2 \oplus \ldots \oplus x_{n - 1}  
 \end{array} \quad
\]

Applying these transformations on the circuit, we get from stage-1 we will 
get a new circuit with up to $3^{n - 1}$ $CNOT$ gates.

\subsection*{\bf Stage 3:}

Now we have a quantum circuit where all $CNOT$ gates are applied on 
the target qubit $t$ with no control qubits with cond-0. 
In stage-3 of our method we carry out minimization to obtain the final
simpler circuit. The method employs the following rule.

\begin{figure} [htbp]
\centerline{\includegraphics[width=11.43cm,height=4.572cm]{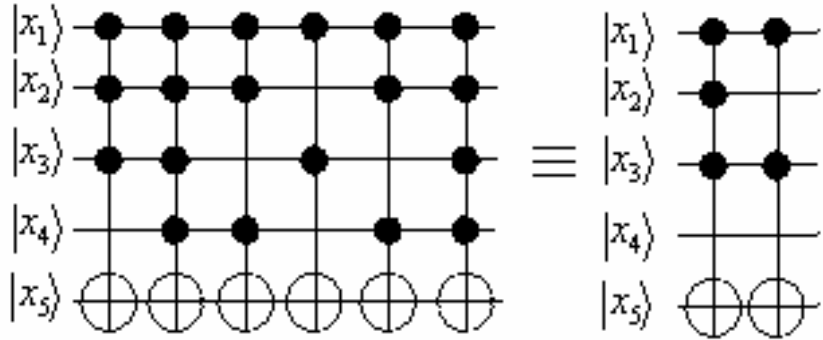}} 
\vspace*{13pt}
\caption{\label{BQCfig12}Rule of minimization.}
\end{figure}

\subsubsection*{Rule of Minimization:}

\begin{equation}
\label{BQCeq16}
\begin{array}{l}
 CNOT(C_i \vert t).CNOT(C_1 \vert t)\ldots CNOT(C_n \vert t).CNOT(C_j \vert 
t) \\ 
 					 = CNOT(C_1 \vert t)\ldots CNOT(C_n \vert t) \\ 
 \end{array} \quad 
\end{equation}

\noindent
if and only if $C_{i}=C_{j}$ as shown in Fig.\ref{BQCfig12}

\begin{figure} [htbp]
\centerline{\includegraphics[width=6.35cm,height=4.318cm]{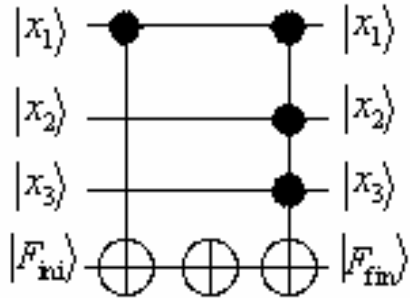}} 
\vspace*{13pt}
\caption{\label{BQCfig13}Final circuit for $F(x_1 ,x_2 ,x_3 ) = \overline {x_1 } + x_2 x_3$.}
\end{figure}

Now applying this rule recursively on the circuit we have, we will get a new quantum circuit 
that on average (see next section) efficiently represents our Boolean function. For 
example, the final quantum circuit for $F(x_1 ,x_2 ,x_3 ) = \overline {x_1 } 
+ x_2 x_3 $ is shown in Fig.\ref{BQCfig13}.

\section{Analysis and Results} 
\begin{figure} [htbp]
\centerline{\includegraphics[width=4.572cm,height=2.5654cm]{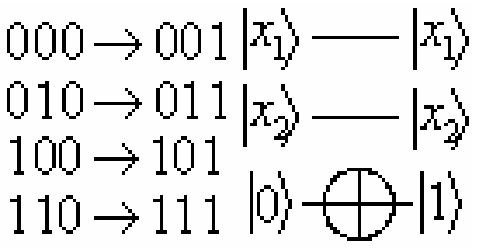}} 
\vspace*{13pt}
\caption{\label{BQCfig14}The most efficient non-zero gate with it's truth table.}
\end{figure}

\noindent
Applying the above method on a truth table, after stage-1 the number of 
gates will be up to $2^{n - 1}$ $CNOT$ gates with some controlled qubits 
with cond-0 and others with cond-1, the case where the number of $CNOT$ gates 
to be $2^{n - 1}$ will occur only when all the $F_{fin}$ being set to 1 in 
the truth table, in this case the most optimum nonzero-gate circuit to be 
found using this method where only $CNOT(F_{ini})$ will exist in the final 
circuit as shown in Fig.\ref{BQCfig14}.

After stage-2, where the transformations being applied to eliminate all $\delta $'s 
equal to 1, we will be left with a new quantum circuit where the number of $CNOT$ gates will be up 
to $3^{n - 1}$ gates. Starting the minimization of the circuit, analysing the 
method for $n$--qubit circuit, we get the following results: the number of possible 
circuits $S$ for all possible input configurations is given by

\begin{equation}
\label{BQCeq17}
S = \sum\limits_{r = 0}^N {\frac{N!}{r!(N - r)!}} = \sum\limits_{r = 0}^N {{ 
}^NC_r } \quad 
\end{equation}

\noindent
where $N = 2^{n - 1}$ and $r$ is the number of $CNOT$ gates to be found in the final 
circuit.

\begin{figure} [htbp]
\centerline{\includegraphics[width=16.75638cm,height=12.98194cm]{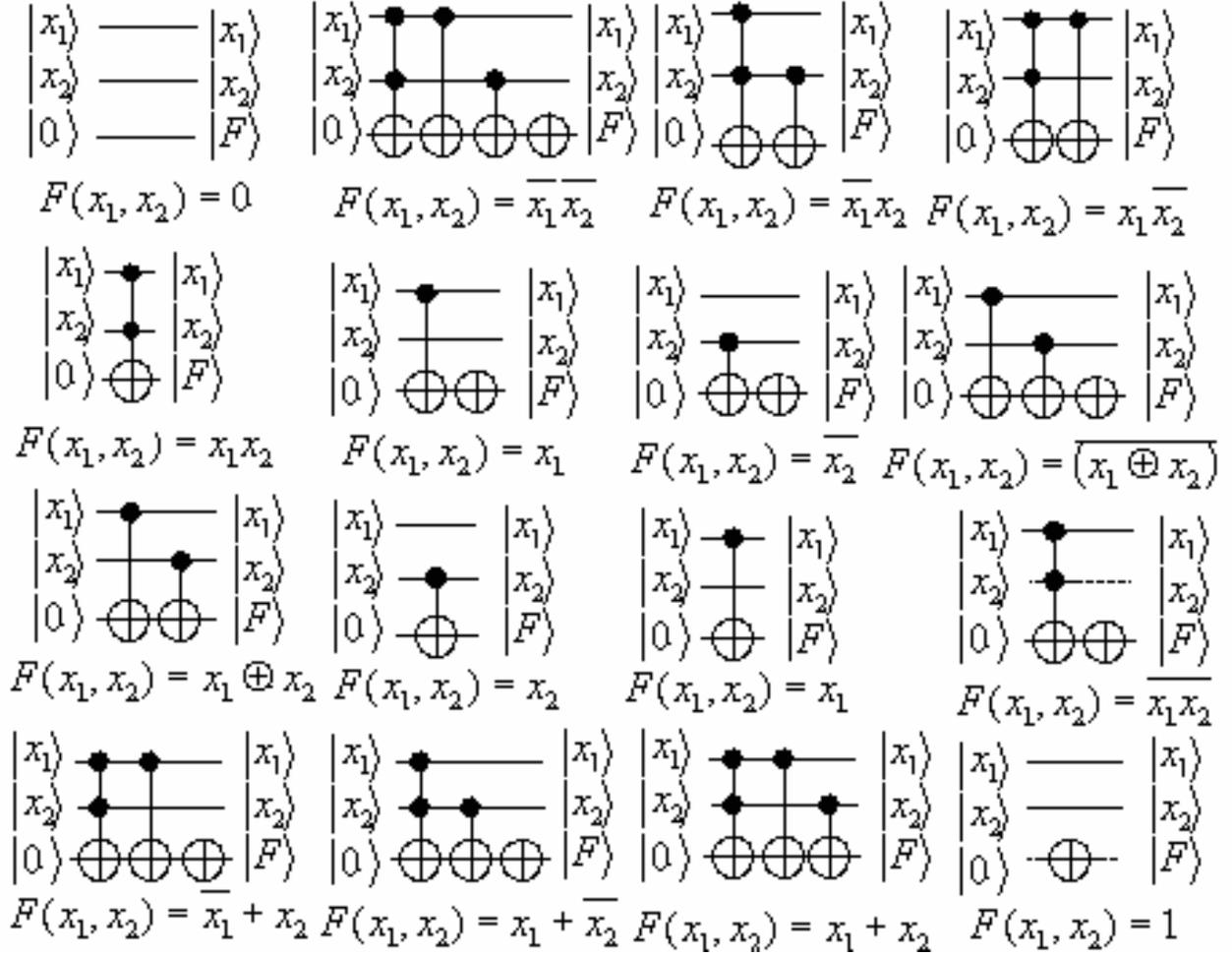}} 
\vspace*{13pt}
\caption{\label{BQCfig3qset}A complete set of a 3--qubit quantum circuits.}
\end{figure}

From Eqn.\ref{BQCeq17} we can see that the probability that the number of $CNOT$ gates in the 
final circuit is $2^{n - 1}$ ($r=N$); which is the worst case, will be $1 / S$ 
which means that the probability for this case will decrease as the number 
of qubits increase, similarly the probability that the final circuit to have 
zero-gates (identity; $r=0$) is $1 / S$ which will decrease as the number of qubits 
increase as well. The most likely case to appear is the \textit{average} where $r = 2^{n - 
2}\left[ {\pm 1} \right]$, for example consider a 3-qubit circuit: The 
number of possible circuits is 16 as shown in Fig.\ref{BQCfig3qset} and the probability 
that a 4-gate circuit or 0-gate circuit to appear is 0.0625 and the 
probability that $2\left[ {\pm 1} \right]$-gates circuit to appear is 0.875. 
And for a 4-qubit circuit: The number of possible circuits is 256 and the 
probability that an 8-gate circuit or 0-gate circuit to appear is 0.00390625 
and the probability that $4\left[ {\pm 1} \right]$-gates circuit to appear 
is 0.7109375.

There is no clear proof at this time that the cost of implementing multiple input
$CNOT$ gates is higher than $CNOT$ gates with fewer inputs, but a further 
optimization for the cost of $CNOT$ gates used in this method can be achieved with a small increase 
in the number of $CNOT$ gates used by applying circuit reduction from the canonical form shown in 
\cite{transrules}.

This method is also used to implement Boolean arithmetic operations like addition and multiplication and also used to 
implement $m$-to-$n$ Boolean logic, these results will be presented in a subsequent paper.

\section{Comparing our Method with Previous Works}
\begin{figure} [htbp]
\centerline{\includegraphics[width=9.24306cm,height=7.830cm]{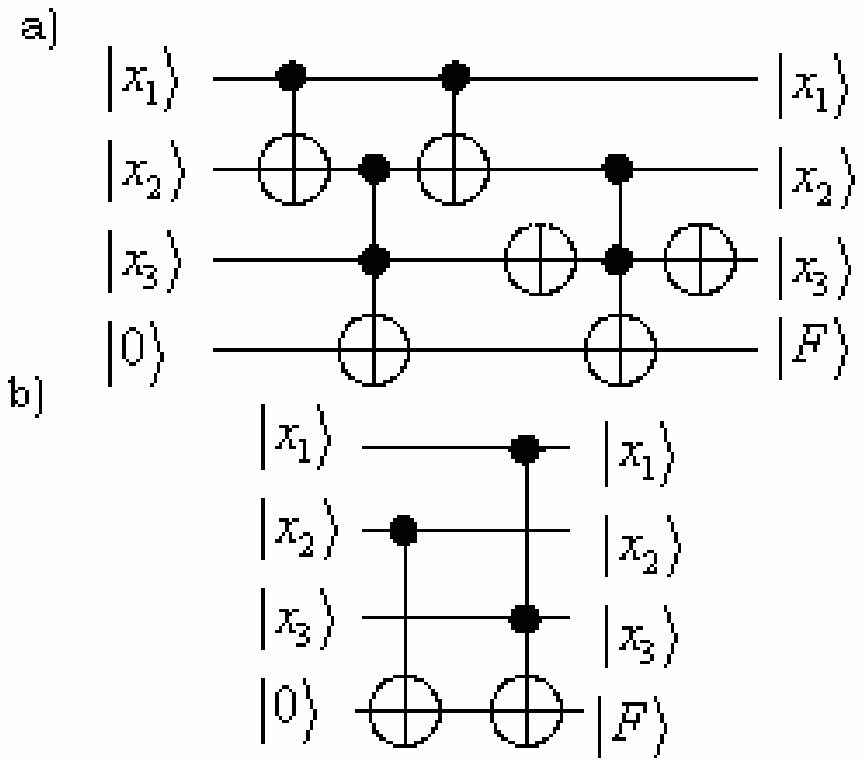}} 
\vspace*{13pt}
\caption{\label{practfig}Implementaion of $F = {\overline {x_1 } x_2  + x_2 \overline {x_3 }  + x_1 \overline {x_2 } x_3 }
$ according to a) method \cite{practmethod}, b) our method.}
\end{figure}

\noindent
The method \cite{practmethod} uses a modified version of {\it Karnaugh maps} and depends on a clever choice of 
certain minterms to be used in minimization process. 
The choice process means that the circuits generated from this method are not unique 
(for example, two alternatives are shown in Fig.\ref{practfig}).Our method will generate a unique form of a 
circuit similar to that shown in Fig.\ref{practfig}(b), which contains a smaller number of $CNOT$ gates. 
The generation of circuits with this method may become very difficult problem for larger quantum circuits 
because of the usage of Karnaugh maps.

\begin{figure} [htbp]
\centerline{\includegraphics[width=14.42974cm,height=10.72642cm]{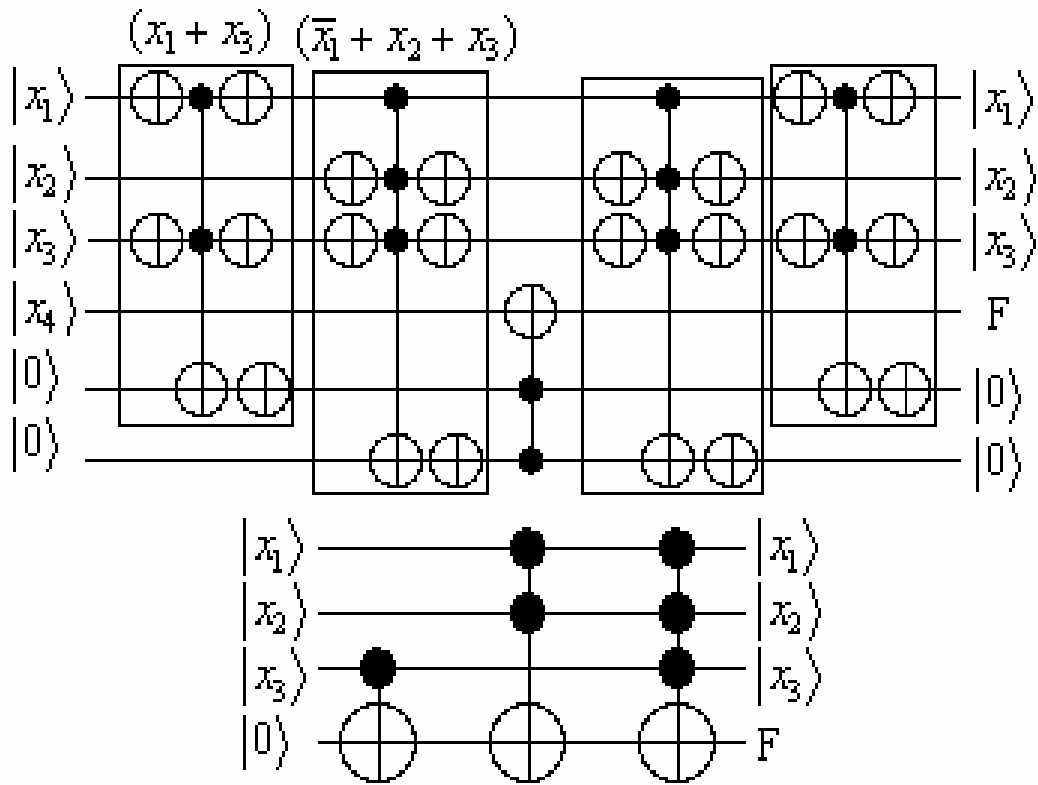}} 
\vspace*{13pt}
\caption{\label{transfig}Implementaion of $F = (x_1  + x_3 )(\overline {x_1 }  + x_2  + x_3 )$ 
according to a) method \cite{transrules}, b) our method.}
\end{figure}

In another work \cite{transrules} a method is described that requires auxiliary qubits in
the quantum circuits obtained. In Fig. \ref{transfig} we give a comparison of a circuit
obtained by this method (Fig. \ref{transfig}(a)) and the equivalent, but much simpler,
circuit obtained by the method proposed in this paper (Fig. \ref{transfig}(b)).

\section{Reversible Version of Classical Operations}
\noindent The construction of reversible classical Boolean circuits has received much attention 
\cite{toffoli80,fredtoff82}. As a direct result from the 3-qubit quantum circuits shown in Fig.15 
, we can pick a set of these circuits which represent the reversible version 
of the classical irreversible operations: AND, OR, NOT, NAND, NOR, XOR and 
XNOR . Using these versions of quantum circuits as reversible classical gates 
together with a reversible FAN-OUT version similar to that shown in \cite{toffoli80} as 
shown in Table.\ref{BQCtab5}, we can build a classical non-quantum reversible version of 
any known digital circuits.

\begin{figure} [htbp]
\centerline{\includegraphics[width=11.1252cm,height=16.5354cm]{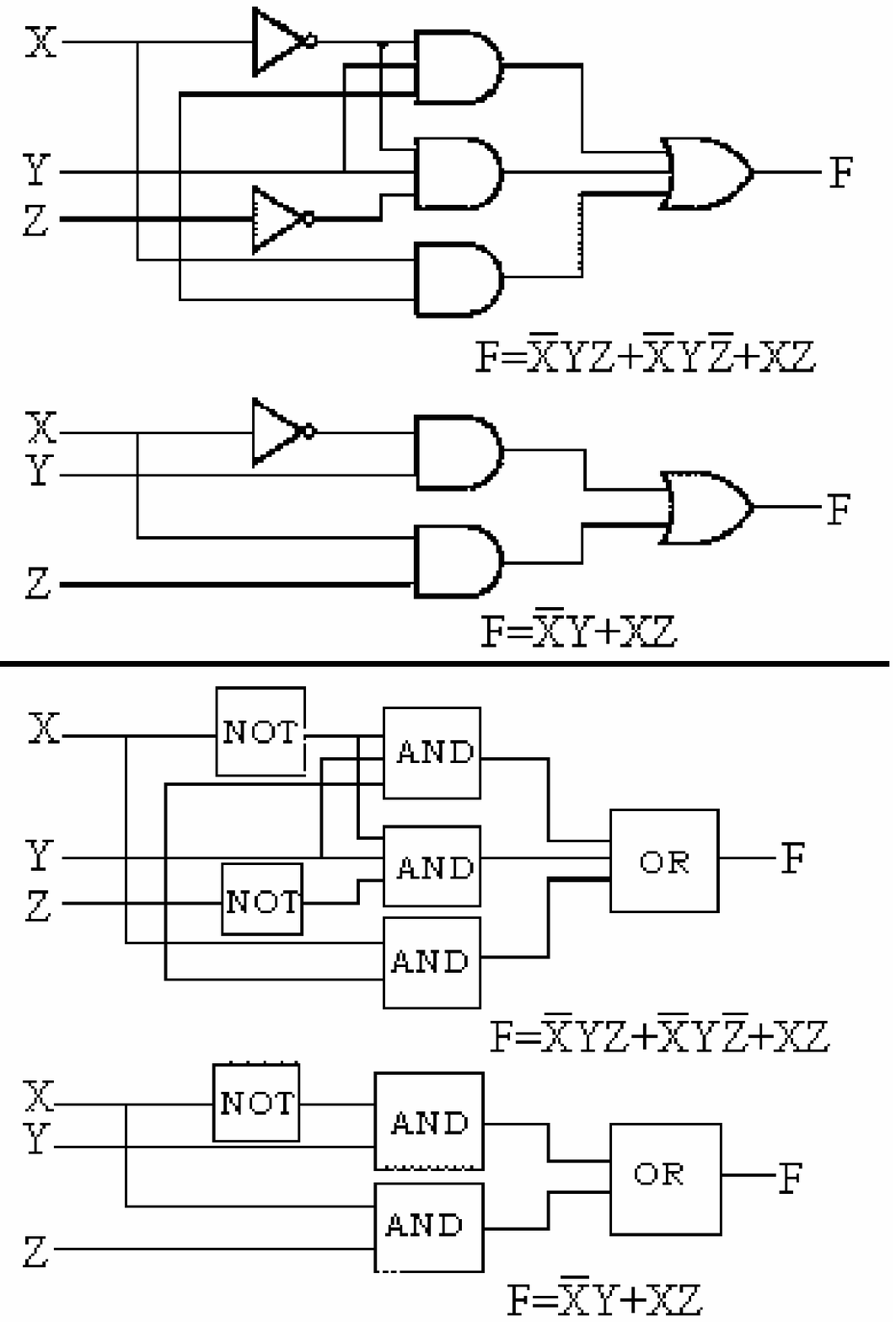}} 
\vspace*{13pt}
\caption{\label{BQCfig39}Converting digital circuits to it's reversible version.}
\end{figure}

Using the gates shown in Table.\ref{BQCtab5}, we can build the reversible version of any 
digital circuit and apply the same methods of simplification and 
optimization applied on those kind of circuits (as shown in Fig.\ref{BQCfig39}). Of course 
this architecture need more investigation to estimate its efficiency.

\begin{table}[H]
\begin{center}
\begin{tabular}{|c|c|c|p{160pt}|}
\hline
Name  & 
$\begin{array}{l}
\mbox{Truth Table \par}   \\ 
x_{1}x_{2}0 \to x_{1}x_{2}F
\end{array}$  & 
Proof  & 
\begin{center}Gate Black-Box\end{center}\\
\hline
AND  & 
$\begin{array}{l}
000 \to 000 \\   
010 \to 010  \\
100 \to 100   \\
110 \to 111  
\end{array}$& 

$\begin{array}{l}
 F = 0 \oplus x_1 x_2 \\ 
 \,\,\,\,\,\, = x_1 x_2 \\ 
 \end{array}$ & 
\begin{center}
\centerline{\includegraphics[width=5.715cm,height=2.54cm]{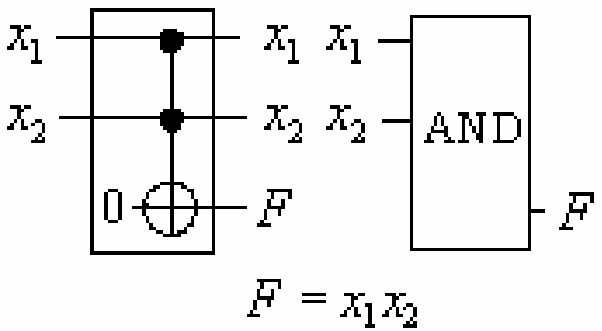}}
\end{center} \par  \\
\hline
OR  & 

$\begin{array}{l}
000 \to 000 \\   
010 \to 011  \\
100 \to 101   \\
110 \to 111  
\end{array}$& 

$\begin{array}{l}
 F = 0 \oplus x_1 x_2 \oplus x_1 \\
 \,\,\,\,\,\,\,\,\,\,\oplus x_2 \\ 
 \,\,\,\,\, = x_1 \overline {x_2 } \oplus \overline {x_2 } \oplus 1 \\ 
 \,\,\,\,\, = \overline {\overline {x_1 } \,\, \overline {x_2 } } \\ 
 \,\,\,\,\, = x_1 + x_2  
 \end{array}$& 
\begin{center}
 \centerline{\includegraphics[width=5.715cm,height=2.54cm]{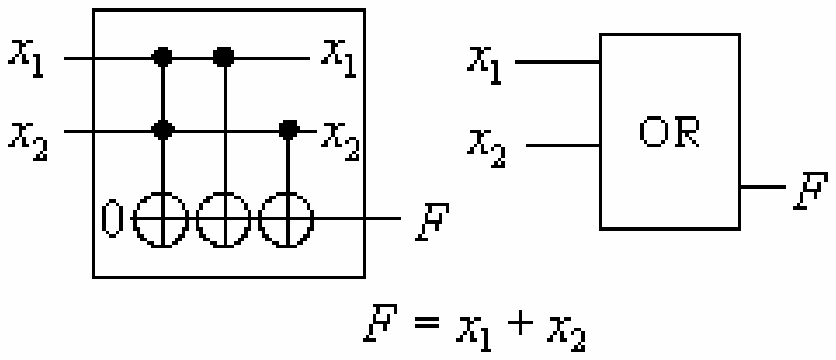}}   
 \end{center}
\par  \\
\hline
NOT  & 
$\begin{array}{l}
0 \to 1 \\   
1 \to 0  
\end{array}$& 

&
\begin{center} 
\centerline{\includegraphics[width=5.715cm,height=1.7018cm]{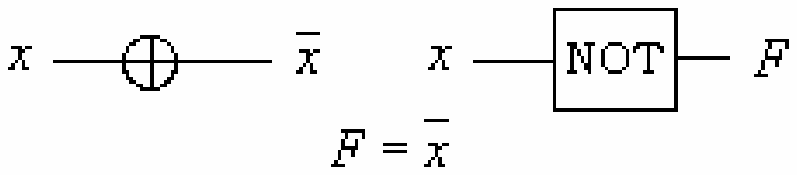}}   
\end{center}
\par  \\
\hline
\end{tabular}
\caption{Reversible version of irreversible classical gates.}
\label{BQCtab5}
\end{center}
\end{table}

\addtocounter{table}{-1}

\begin{table}[H]
\begin{center}
\begin{tabular}{|c|c|c|p{120pt}|}
\hline
Name  & 
$\begin{array}{l}
\mbox{Truth Table \par}   \\ 
x_{1}x_{2}0 \to x_{1}x_{2}F
\end{array}$  & 
Proof  & 
\begin{center}Gate Black-Box\end{center}\\
\hline
NAND  & 
$\begin{array}{l}
000 \to 001 \\   
010 \to 011  \\
100 \to 101   \\
110 \to 110  
\end{array}$& 
$\begin{array}{l}
 F = \overline {0 \oplus x_1 x_2 } \\ 
 \,\,\,\,\,\, = \overline {x_1 x_2 }  
 \end{array}$& 
\begin{center}
 \centerline{\includegraphics[width=3.715cm,height=2.2718cm]{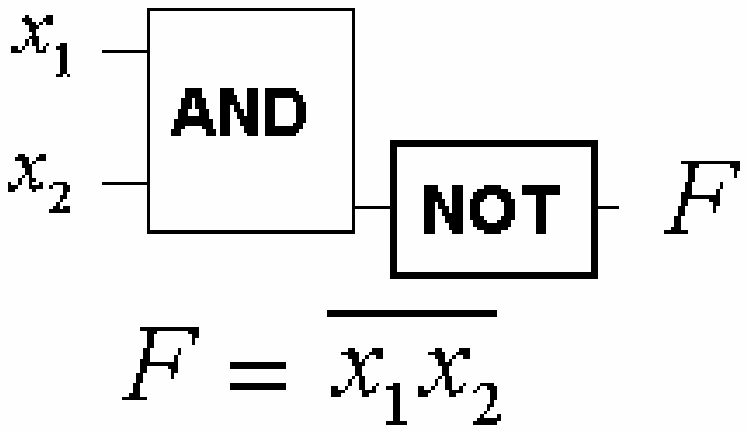}}   
 \end{center}
\par \newpage  \\
\hline 
NOR  & 
$\begin{array}{l}
000 \to 001 \\   
010 \to 010  \\
100 \to 100   \\
110 \to 110  
\end{array}$& 
$\begin{array}{l}
 F = \overline {0 \oplus x_1 x_2 \oplus x_1 } \\
 \,\,\,\,\,\,\,\, \overline {\oplus x_2 } \\ 
 \,\,\,\,\, = \overline {x_1 \overline {x_2 } \oplus \overline {x_2 } \oplus 1} \\ 
 \,\,\,\,\, = \overline {x_1 } \overline {x_2 } \\ 
 \,\,\,\,\, = \overline {x_1 + x_2 }  
 \end{array}$& 
\begin{center}
 \centerline{\includegraphics[width=3.715cm,height=1.5718cm]{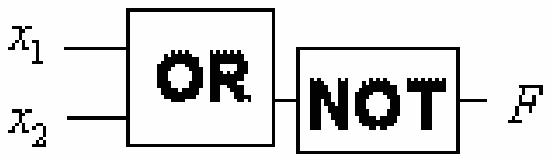}}   
 \end{center}
\par  \\
\hline
FAN-OUT  & 
$\begin{array}{l}
00 \to 00 \\   
10 \to 11   
  
\end{array}$& 

$0 \oplus x = x$& 
\begin{center}
\centerline{\includegraphics[width=3.715cm,height=1.8772cm]{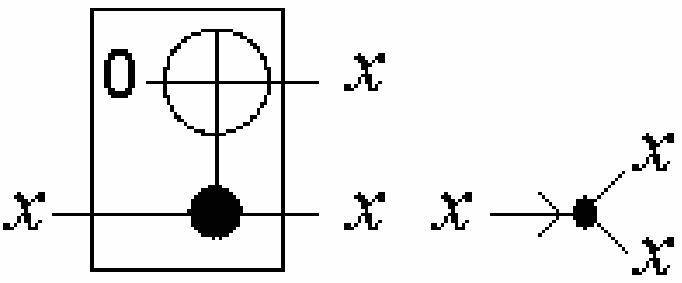}}   
\end{center}
\par  \\
\hline
\end{tabular}
\caption{Continued...}
\label{BQCtab5_1}
\end{center}
\end{table}

\addtocounter{table}{-1}

\begin{table}[H]
\begin{center}
\begin{tabular}{|c|c|c|p{163pt}|}
\hline
Name  & 
$\begin{array}{l}
\mbox{Truth Table \par}   \\ 
x_{1}x_{2}0 \to x_{1}x_{2}F
\end{array}$  & 
Proof  & 
\begin{center}Gate Black-Box\end{center}\\
\hline

XOR  & 
$\begin{array}{l}
000 \to 000 \\   
010 \to 011  \\
100 \to 101   \\
110 \to 110  
\end{array}$& 
 
$\begin{array}{l}
 0 \oplus x_1 \oplus x_2 \\ 
 = x_1 \oplus x_2  
 \end{array}$& 
 \begin{center}
\centerline{\includegraphics[width=6.096cm,height=2.9972cm]{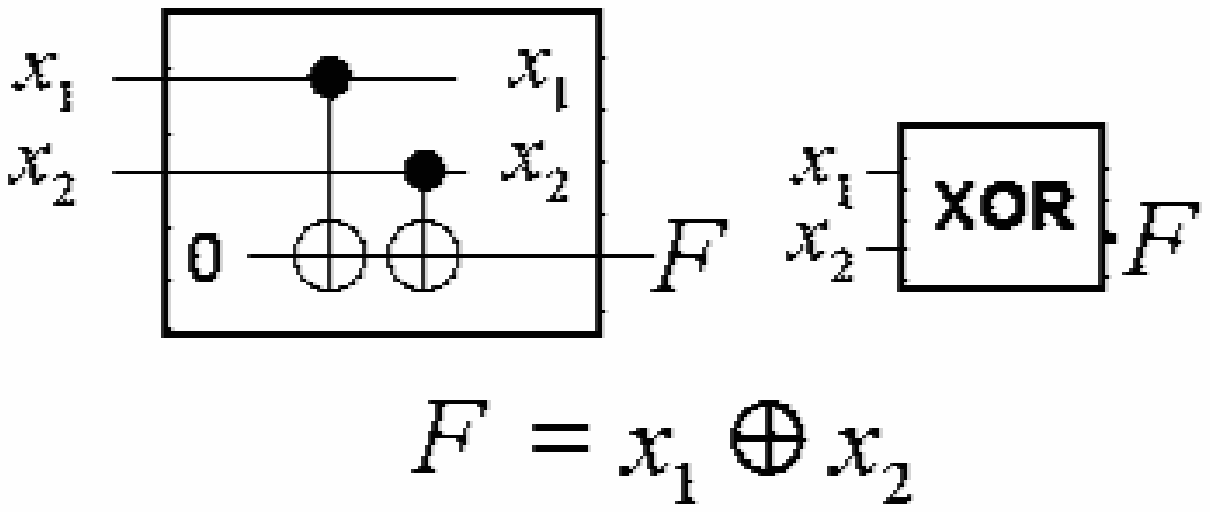}}  
\end{center}
 \par  \\
\hline
XNOR  & 
$\begin{array}{l}
000 \to 001 \\   
010 \to 010  \\
100 \to 100   \\
110 \to 111  
\end{array}$& 

$\begin{array}{l}
 \overline {0 \oplus x_1 \oplus x_2 } \\ 
 = \overline {x_1 \oplus x_2 } \\ 
 \end{array}$& 
\begin{center}
 \centerline{\includegraphics[width=5.8452cm,height=3.02252cm]{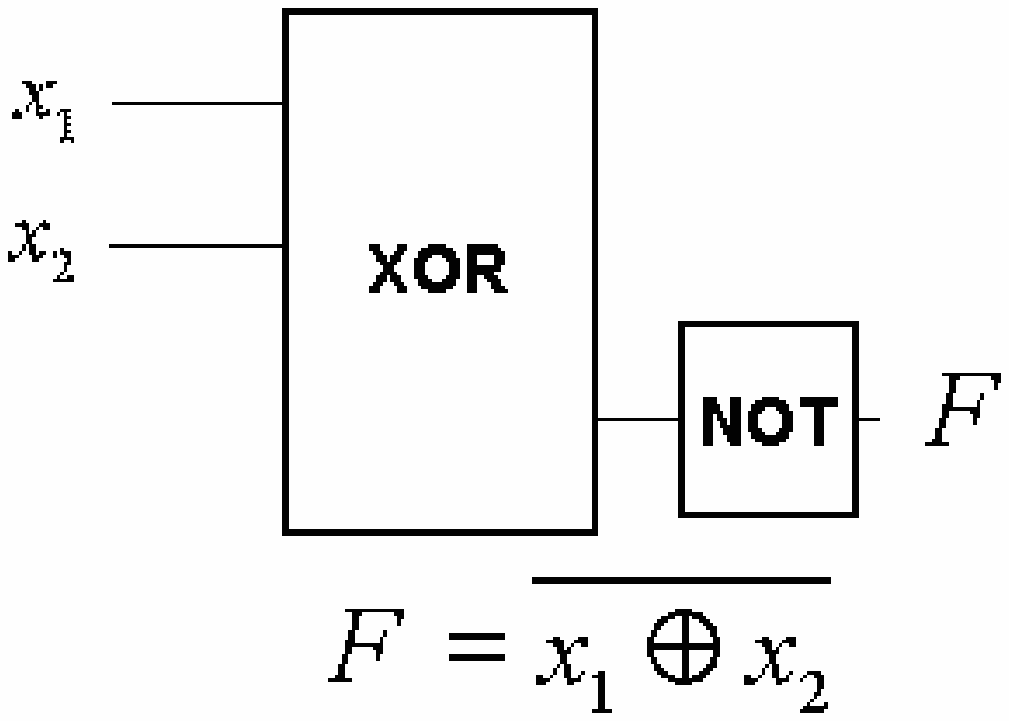}}   
 \end{center}
\par  \\
\hline
\end{tabular}
\caption{Continued...}
\end{center}
\end{table}

\end{document}